\documentstyle[psfig]{lamuphys}
\makeatletter
\let\chapter\hid@chapter
\makeatother

\begin{document}
\pagenumbering{arabic}

\title{The Hot Galactic Corona and the Soft X-ray Background}

\author{
Q. Daniel Wang
}

\institute{
Dearborn Observatory, Northwestern University
2131 Sheridan Road, Evanston,~IL 60208-2900, USA
}

\maketitle

%
%
\newcommand{\R}{~ROSAT}
\newcommand{\RAS}{\R all sky survey}
\newcommand{\ROSAT}{{\it ROSAT }}
\newcommand{\IRAS}{{\it IRAS }}
\newcommand{\COBE}{{\it COBE }}
\newcommand{\RASS}{{\it ROSAT} All-Sky Survey }
\newcommand{\EXSAS}{{\it EXSAS }}
\newcommand{\HEAO}{{\it HEAO-1 }}
\newcommand{\etal}{et~al.\,}
\newcommand{\SEC}{^s\!\!.}
\newcommand{\DEG}{^\circ}
\newcommand{\AMIN}{^\prime}
\newcommand{\avrg}[1]{\langle{#1}\rangle}
\def\gsim{\lower 2pt \hbox{$\, \buildrel {\scriptstyle >}\over
{\scriptstyle \sim}\,$}}
\def\lsim{\lower 2pt \hbox{$\, \buildrel {\scriptstyle <}\over
{\scriptstyle \sim}\,$}}

\begin{abstract}
I characterize the {\sl global} distribution of the 3/4~keV band
background with a simple model of the hot Galactic corona, plus an 
isotropic extragalactic background. The corona is assumed to
be approximately polytropic (index = 5/3) and hydrostatic in the 
gravitational potential of the Galaxy. The model accounts for X-ray 
absorption, and is constrained iteratively with the \ROSAT all-sky X-ray 
survey data. Regions where the data deviate significantly from the model
represent predominantly the Galactic disk and individual nearby 
hot superbubbles. The global distribution of the background, outside 
these regions, is well characterized by the model; the $1\sigma$ relative 
dispersion of the data from the model is $\sim 15\%$. The electron density 
and temperature of the corona near the Sun are $\sim 1.1 \times 10^{-3} 
{\rm~cm^{-3}}$ and $\sim 1.7 \times 10^6$~K. The same model also explains 
well the 1.5~keV band background. The model prediction in the 1/4~keV band, 
though largely uncertain,  qualitatively 
shows large intensity and spectral variations of the corona
contribution across the sky.

\end{abstract}

\section{Introduction}

	Although Spitzer (1956) speculated the presence of a hot 
($\sim 10^6$~K) Galactic corona around the Milky Way more than 40 years ago, 
direct observational evidence of it comes only recently. The very existence 
of hot gas far away from the Sun is shown convincingly first by the
the detection of the shadow cast by the Draco cloud ($D \gsim 300$~pc; 
$z \gsim 200$~pc) against the 1/4~keV background
(Burrows \& Mendenhall 1991; Snowden et al. 1991).
Based on the X-ray shadowing of the Magellanic Bridge ($D \sim 60$~kpc), 
Wang \& Ye (1996) further find that $\sim 30\%$ of the background observed at
$\sim 0.7$~keV is Galactic in origin. At this energy, however, the 
contribution from the Local Hot Bubble around the Sun appears negligible
(Snowden, McCammon, \& Verter 1993; Kuntz, Snowden, \& Verter 1997), 
although the Bubble is responsible for most of the background in the 1/4~keV 
band (e.g., Snowden et al. 1997a). Moreover, the \ROSAT all-sky survey 
(Snowden et al. 1995; 1997b) clearly shows an overall intensity
enhancement in the 0.5-2~keV range over the Galactic center hemisphere. 
Snowden et al. have suggested that this enhancement 
may represent a bulge of hot gas around the Galactic center. 

	I have tested how a simple, physically self-consistent model of the 
hot Galactic corona may account for the soft X-ray background, especially in 
the 3/4~keV band. The radiation in this band is sensitive to gas at 
temperatures $\gsim 1.5 \times 10^6$~K and is not as heavily attenuated by the 
X-ray-absorbing interstellar medium as in the 1/4~keV band. Furthermore,
the effect of the small-scale ($\lsim 1^\circ$) clumpiness of the 
medium may be negligible over a unit X-ray
absorption depth of $\sim 2 \times  10^{21} {\rm~cm^{-2}}$.
It is clear, though, that the observed background is strongly contaminated 
by various discrete X-ray sources in regions near the Galactic 
plane and by a few high 
Galactic latitude features such as Loop I and Eridanus superbubbles. I have
therefore devised data reduction and analysis algorithms to minimize the 
effects of these contaminations. The results are encouraging. The model 
provides a frame work for both characterizing the global hot gas distribution 
of the Galactic corona and defining discrete X-ray-emitting features.

	In the following, I first talk about the data analysis and 
modeling procedures, and then present some preliminary results. Finally, 
I make some comparisons of the results with other independent 
measurements and extend the results into the 1/4~keV and 1.5~keV bands.

\section{Data Analysis}

	The data used in this analysis come from the all-sky 
surveys made with \ROSAT (e.g., Snowden et al. 1995) and \IRAS 
(e.g., Boulanger \& Perault 1988). The \ROSAT data, in three energy
bands ($\sim 1/4$~keV, 3/4~keV, 1.5~keV), were 
down-loaded from MPE (http://www.rosat.mpe-garching.mpg.de/),
and the \IRAS 100$\mu$m\ data from HEASARC (http://skview.gsfc.nasa.gov/).
The data were presented as surface brightness intensity maps 
(480$\times$240 pixels),
Aitoff-projected in the Galactic coordinates. These maps, though not reflecting
the intrinsic spatial resolutions (a few arcminutes) of the surveys, 
provide a convenient database for investigating the global 
distributions of both the X-ray background and the X-ray-absorbing medium. 

	For my data modeling purpose, I processed the data in three major 
steps. First, to approximately correct for the zero intensity level of 
the \IRAS data, I shifted the 100$\mu$m\ map to match the estimated total 
absorption $9 \times 10^{19} {\rm~cm^{-2}}$ in the Lockman Hole region
(Snowden et al. 1994), using a conversion of 1.14 $\times 10^{20} 
{\rm~cm^{-2}/(MJy~sr^{-1}})$ (Boulanger \& Perault 1988; 
Snowden et al. 1997a). 
Second, I removed strips of pixels that were not, or very poorly, covered 
in the surveys. Third, I compressed the maps by taking a median summary in 
each {\sl non-overlapping} square of 3$\times$3 pixels. I discarded
any square that contained one or more removed pixels (including those 
outside the sky projection). The median summary, as a resistant statistic, 
effectively removed the effects of outstanding small-scale ($\lsim 1$~deg) 
features (e.g., nearby galaxies and AGNs). The top panels of Fig. 1 
present the processed data.

\section{Modeling}

\begin{figure}[thbp]
\psfig{file=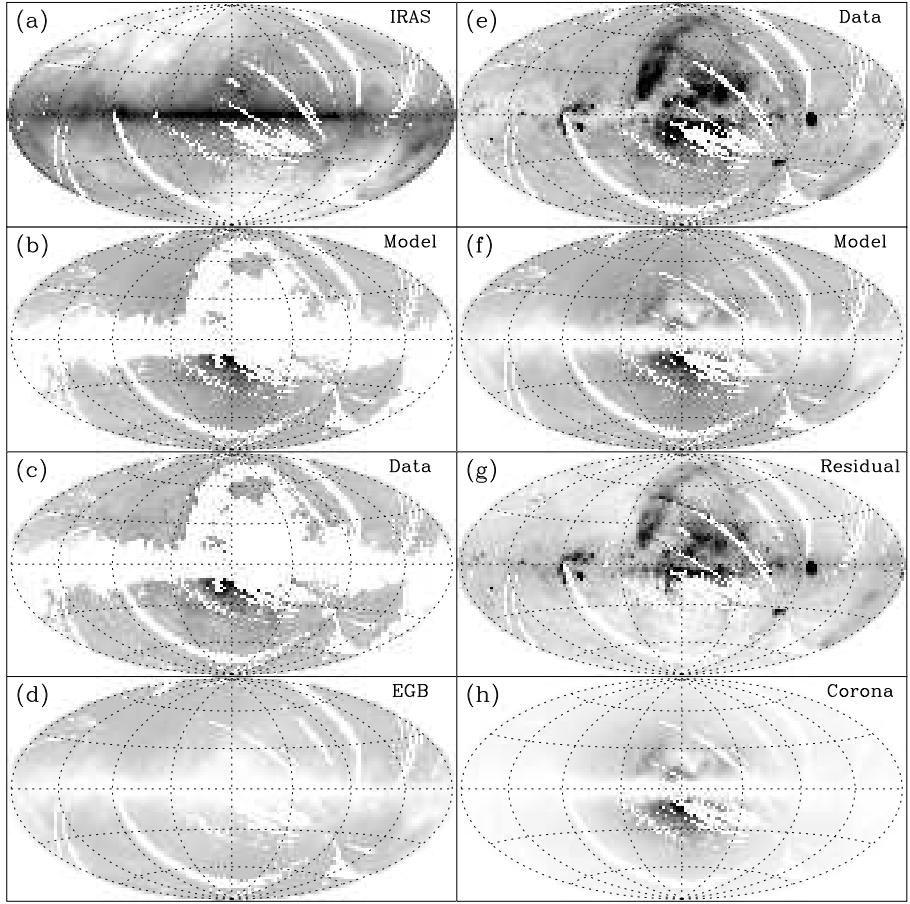,width=\hsize,clip=}
\caption[]{Surface brightness intensity maps in the zero-centered Galactic 
coordinates with Aitoff equal-area projections. Both X-ray data and model 
are in the 3/4~keV band. From the top panel to the bottom one in the left 
column are the zero-level corrected \IRAS 100$\mu$m\ survey (a); Model map
covering only pixels used in the fit at the threshold
0.4 (b); \ROSAT data in the same region  (c); and the extragalactic 
background component 
of the model (d). The panels in the right column, arranged for an easy
comparison with Fig. 4, are the \ROSAT data (e); the model including 
regions not used in the fit (f); the residual of the \ROSAT survey minus 
the model (g); the corona component of the model (h). 
The false color range is between
1 and $440{\rm~MJy~sr^{-1}}$ logrithmetically for (a), 0 to 5 $\times 10^{-4} 
{\rm~counts~s^{-1}~arcmin^{-2}}$ linearly for (b), (c), (d), (e), (f), and 
(h), and -0.7 to 5 $\times 10^{-4} {\rm~counts~s^{-1}~arcmin^{-2}}$ for (g); 
the black represents the lowest intensity while the yellow is the highest. 
(Referee: The color version of this figure can be found at
http://www.astro.nwu.edu/astro/wqd/paper/halo/)}
\label{h72_fig01}
\end{figure}

	To capture the global distribution of the X-ray background, 
I have considered a model that consists of two contributions: 
an isotropic extragalactic background (EGB) and an axisymmetric 
Galactic corona. The corona
is assumed to be quasi-hydrostatic in the gravitational 
potential of the Galaxy (Wolfire et al. 1995; Johnston, Spergel, \& Hernquist
1995). The potential is a sum of three components: a Miyamoto-Nagai disk,
a spheroid bulge, and a logarithmic halo. These components together
provide a nearly flat rotation
curve from 1 to 30~kpc with a circular velocity of $225 {\rm~km~s^{-1}}$
at the Galacto-centric radius 8.5~kpc of the Sun. It is the disk component,
however, that is chiefly responsible for the corona structure visible in 
the \ROSAT band. For ease of modeling, I neglect the angular momentum of 
the corona. The momentum should be small if the corona is fueled primarily 
by hot gas from the Galactic central region.  Observations of nearby disk 
galaxies do indicate that hot gas outflows happen primarily in galactic
center regions (e.g., Wang et al. 1995; Pietch, Supper, \& Vogler 1995). 
Furthermore, because
the cooling of the hot gas at temperatures $\gsim 10^6$~K is most likely 
adiabatic, a polytropic equation of state of index 5/3 may be a reasonably
good description of the hot gas. The shape and normalization of this
model corona are determined by two adjustable parameters, chosen here 
to be the electron density ($n_{o}$) and temperature ($T_{o}$)
of the corona near the Sun.

	I calculate a model X-ray background, accounting for 
both the X-ray absorption and the {\sl ROSAT}/PSPC spectral response.  The
calculation  makes the following assumptions: (1) The EGB has a spectrum 
characterized by a power law of energy slope equal to 1 
(e.g., Hasinger et al. 1993); (2) The hot gas of the corona is in a collisional
ionization equilibrium (Raymond \& Smith 1977); (3) Both the hot gas and the 
X-ray-absorbing medium are of solar metal abundances; (4) The absorption 
is foreground, which should be reasonably good except for regions close to
the Galactic plane ($|b| \lsim 10^\circ$). X-ray 
surface brightness intensities of a unit volume are calculated first in a grid
as a function of hot gas temperature and X-ray-absorbing medium column 
density. Based on this grid, an intensity integration along a line of sight 
can be carried out efficiently. The integrated intensity of the corona, 
together with a partially absorbed EGB contribution, constitutes the model 
intensity $m_i$, where $i$ represents a line of sight, or a pixel in a 
background map (Fig. 1).

The intrinsic intensity $I_e$ of the EGB as well as the corona parameters, 
$n_{o}$ and $T_{o}$, are constrained by
a model fit to the \ROSAT survey data $d_i$ in the 3/4~keV band. 
Specifically, the fit minimizes the statistic 
$\chi^2 = \sum [(d_i-m_i)w_i]^2$, where the summation is over the adopted map 
pixels. The weight $w_i$ is 
inversely proportional to the pixel-to-pixel intensity dispersion, 
$\sim 15\% m_i$, estimated 
in regions away from local X-ray enhancements (see also \S 4). 
In comparison, the counting statistical uncertainty is considerably
smaller ($\lsim 5\%$) in a $\sim 4~{\rm deg}^2$ pixel 
with a typical exposure $\sim 500$~s of the \ROSAT survey. 
Adding this uncertainty in the weight would cause only minor ($\lsim 5\%$) 
changes in the results. The fit is conducted iteratively to remove 
outliers --- pixels where the data deviate significantly from the model.
Starting with $w_i = $ constant, the fit gets an initial estimate of the 
model $m_i$. $w_i$ is then recalculated. After removing those 
pixels with relative deviations $(d_i-m_i)w_i > 2$, an arbitrarily chosen 
initial threshold, the fit proceeds to minimize 
$\chi^2$ again and so on, until no more pixel has 
$(d_i-m_i)w_i > 2$. Next, the threshold is reduced by 10\%, 
an arbitrarily chosen step, and a new round of fitting and 
pixel removing starts. Fig. 2 shows how the model parameter fits evolve 
with the deviation threshold for pixel-removing. When the threshold is 
large, the outliers strongly influence the fit of the parameters, especially
$n_o$ and $I_e$. The parameter values also fluctuate until 
the threshold is reduced to $\sim 0.2$, when nearly half of the map pixels
are removed. The results are not particularly sensitive to
either the initial threshold or the step.

\begin{figure}[thbp]
\psfig{file=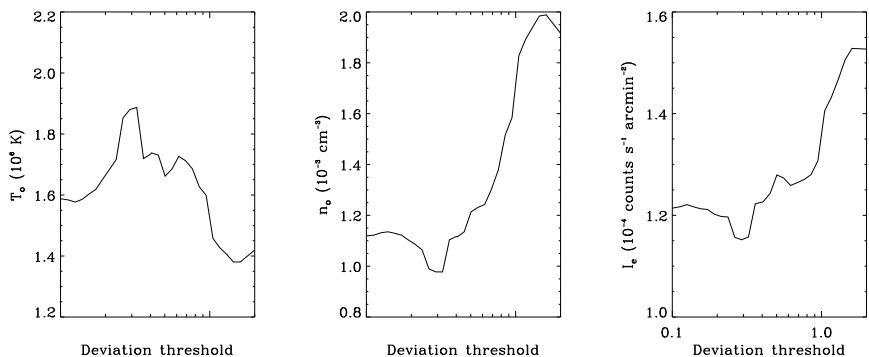,width=\hsize,clip=}
\caption[]{Parameter changes with the pixel-removing threshold. 
The model fitting proceeds from the right to the left.}
\label{h72_fig02}
\end{figure}

\section{Results}

	Fig 1 includes a comparison of the data with the model 
background fitted at the threshold equal to 0.4, at which the parameter 
values are close to the converging values at $\lsim 0.2$ and only a 
quarter of the pixels are removed. While the model is {\sl intrinsically} 
symmetric relative to the Galaxy 
rotation axis, the asymmetric appearance of the model background in Fig. 1
(e.g., Panel f) is due to the foreground X-ray absorption (a).
The dispersion of the data from the model is $\sim 0.15\%$, averaged over the 
pixels remained in the fit (Panels b and c in Fig. 1). 
Pixels excluded from the fit are almost entirely at 
the Galactic plane and in high Galactic latitude regions that are
contaminated by Loop I (including the Sco-Cen starforming region) and 
Eridanus superbubbles as well as the Large 
Magellanic Cloud near the south Elliptical pole (see Snowden et al. 1995
for a graphic illustration). These features and the disk component
stand out in the residual map, because of the removal of 
global emission and absorption effects of the corona and EGB 
contributions. Part of the residual emission near the Galactic plane, 
however, may result from an incomplete
removal of the corona contribution. The foreground assumption of the 
X-ray-absorbing medium leads to a slight overestimation of the absorption 
near the Galactic plane, where some of the medium is located behind the 
corona emission. While the corona contribution (h) is 
strongly concentrated in the Galactic center hemisphere, the EGB (d) 
is more uniformly distributed in the sky, except for 
regions close to the Galactic plane.

	Since the model describes the global distribution of the 3/4~keV
background reasonably well, it is tempting to use the model fit as a 
characterization of the hot Galactic corona. From Fig. 2, I estimate
the local hot gas electron density and temperature of the corona
as $\sim 1.1 \times 10^{-3} 
{\rm~cm^{-3}}$ and $\sim 1.7 \times 10^6$~K. The uncertainties in these 
two parameters are $\sim 10\%$, corresponding to
the fluctuations of the parameter values  
in the threshold range of $\lsim 0.5$.
Fig. 3 shows the spatial distribution of hot gas temperature
in the corona. The density and temperature at the Galactic center
are  $\sim 1.2 \times 10^{-2} {\rm~cm^{-3}}$ and $\sim 8.5 \times 10^6$~K.
The integrated mass, thermal energy, and bolometric luminosity of gas at 
temperatures $> 10^6$~K are $3 \times 10^7 M_\odot$, $\sim 
2 \times 10^{56}$~ergs, and $\sim 2 \times 10^{40} {\rm~ergs~s^{-1}}$
(only 15\% in the 0.5-2~keV range), respectively. The inferred 
mean radiative cooling timescale of the gas is then $\sim 
3 \times 10^8$~years. The radiative cooling accounts for about 2\% of the total
mechanical energy input from supernovae in the Galaxy (assuming  
one supernova per 30~years and $10^{51} {\rm~ergs}$ per supernova).

\begin{figure}[thbp]
\psfig{file=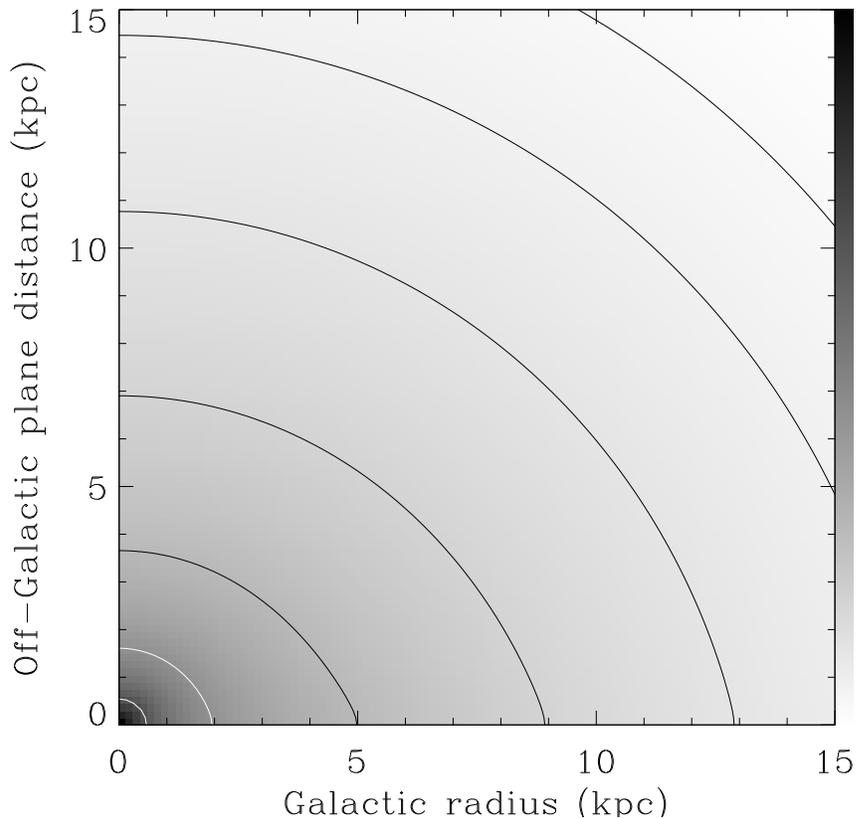,width=\hsize,clip=}
\caption[]{Model gas temperature distribution of the corona in 
the Galactic cylindrical coordinates. The gray-scale is in the range of 
0.18 to 8.4 $\times 10^6$~K,
while the contours are at 0.5, 0.75, 1.1, 1.7, 2.5, 3.8, and 5.7 
$\times 10^6$~K.}
\label{h72_fig03}
\end{figure}

\section{Discussion}

	The above corona model, constrained by the spatial distribution 
of the 3/4~keV band background, makes specific predictions that can be 
compared with various independent measurements.  First, 
the model predicts a thermal pressure of the corona near the Sun as 
$P/k \sim 3.6 \times 10^3 {\rm~K~cm^{-3}}$. This prediction is within
the uncertainty range of $\sim 2 \times 10^3 - 4 \times 10^3
{\rm~K~cm^{-3}}$ inferred from CIV emission/absorption
lines (Martin \& Bowyer 1990; Shull \& Slavin 1994) and 
from the two-phase structure of high-velocity clouds (Wolfire et al. 1995).
Second, Fig. 3 shows that the spectral characteristics of coronal gas 
varies strongly in the sky. Based on broad-band spectral properties of
the soft X-ray background, various attempts have been made to 
estimate the average temperature of hot gas beyond the Local Bubble 
(e.g., Garmire et al. 1992; Wang \& McCray 1993; Kerp 1994; Sidher et al. 
1996; Snowden et al. 1997b). Such an estimate depends on an assumption 
about the EGB spectrum, which remains poorly constrained 
in the 0.1-2~keV range. Nevertheless, the estimated 
hot gas temperature falls within a range between $\sim 10^{6.0}$~K 
and $\sim 10^{6.4}$~K. This range can in principle be reproduced with the 
model, depending on specific lines of sight (Fig. 3). Third, 
the model predicts a corona contribution of $\sim 3 \times 10^{-5} 
{\rm~counts~s^{-1}~arcmin^{-2}}$ in the PSPC R4 band, which is centered 
around $0.7$~keV (Snowden et al. 1997b),
toward the X-ray shadowing cloud ($l, b = 295^\circ, -42^\circ$) in 
the Magellanic Bridge. This contribution is consistent with
the measured {\sl Galactic} component of $2.7 (1.5-3.7) \times 10^{-5} 
{\rm~counts~s^{-1}~arcmin^{-2}}$ (90\% confidence interval; Wang \& Ye 1996).
Thus the corona model is useful for a uniform explanation of various 
observations of distant hot gas at high Galactic latitudes.

\begin{figure}[thbp]
\psfig{file=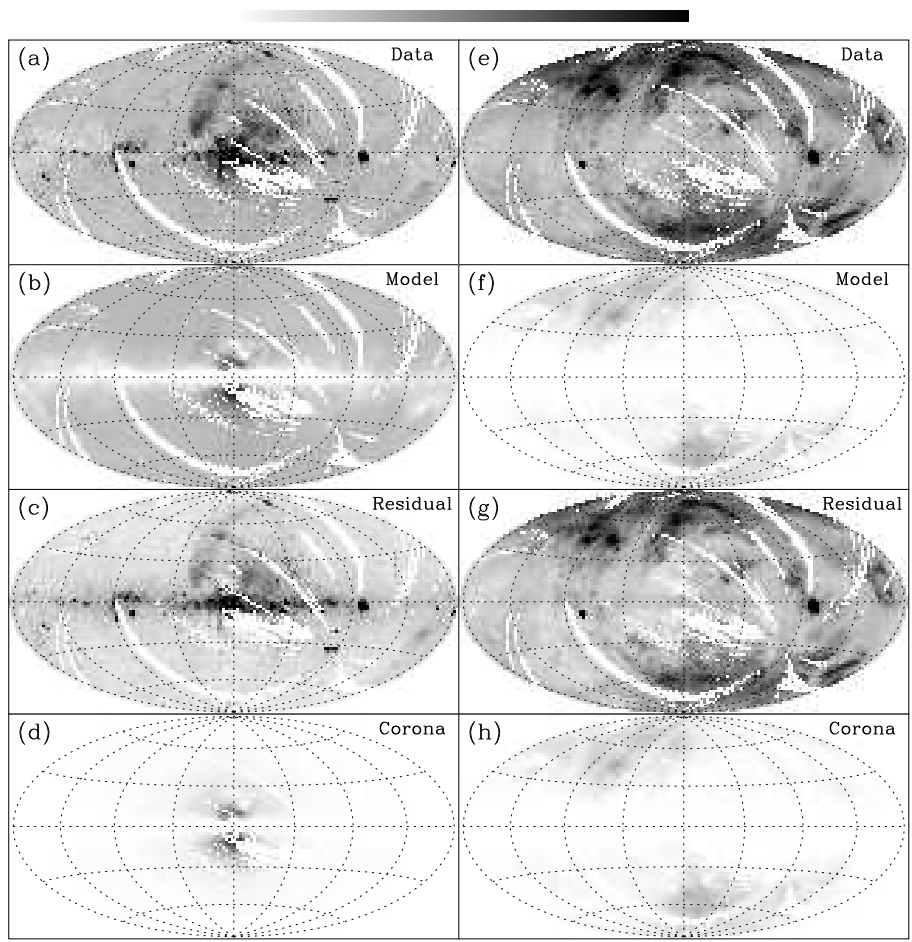,width=\hsize,clip=}
\caption[]{Data and model predictions in the 1.5~keV band (left
column) and in the 1/4~keV band (right column). The projections and sky 
coverage are the same as those in Fig. 1.
From the top row to the bottom one are the \ROSAT 
survey data, the model predictions of the corona plus the extragalactic 
background, the residual maps of the survey data minus the model,
and the corona components of the model in the two bands. The false color range
in the left column spans between 0 to 5 $\times 10^{-4} 
{\rm~counts~s^{-1}~arcmin^{-2}}$ for the panels (a), (b) and (d) and -0.7 to 5 
$\times 10^{-4} {\rm~counts~s^{-1}~arcmin^{-2}}$ for (c); the range
in the right column is all between 0 to 15 $\times 10^{-4} 
{\rm~counts~s^{-1}~arcmin^{-2}}$. 
(Referee: The color version of this  figure can be 
found at  http://www.astro.nwu.edu/astro/wqd/paper/halo/)}
\label{h72_fig04}
\end{figure}

	Fig. 4 further compares the model {\sl predictions} in the 1.5~keV and 
1/4~keV bands with the corresponding \ROSAT survey data. The data in these 
bands are processed in the same way as in the 3/4~keV band (\S 2). The 
intrinsic EGB intensity is fixed as 1.4 $\times 10^{-4} 
{\rm~counts~s^{-1}~arcmin^{-2}}$ in the 1.5~keV band, accounting for the total
background observed in high Galactic latitude, anti-Galactic center 
hemisphere, and as 4.2 $\times 10^{-4} 
{\rm~counts~s^{-1}~arcmin^{-2}}$ in the 1/4~keV band 
(Barber, Roberts, \&  Warwick 1996). The model
explains well the global distribution of the background in 1.5~keV band.
Residual features (Panel c) morphologically mimic those in the 3/4~keV band. 
The corona contribution is confined more into the central region of 
the Galaxy than in the 3/4~keV band. 

In the 1/4~keV band, the model demonstrates that 
the corona contribution may vary significantly across the 
sky in both intensity and spectrum. The model predicts that up to $\sim 45\%$ 
of the observed background intensity may arise in the corona, depending on
individual lines of sight.  In the direction of the Draco cloud 
($l, b \sim 90^\circ, 39^\circ$), for example, the corona contribution
is about 30\%, which is significantly less than the distant
component ($\sim 60\%$) inferred from the X-ray shadowing measurement 
(e.g., Burrows \& Mendenhall 1991). The extragalactic component can
account for an additional $\sim 10\%$. The rest of the distant component 
may arise in a disk component that is not included in the model.
In addition,  the model predication may be an underestimation of
the true corona contribution in the 1/4~keV band. The assumed ionization 
equilibrium may break down for gas of $< 10^6$~K, because the radiative 
cooling timescale becomes shorter than the relevant recombination timescales 
(e.g., Breitschwerdt \& Schmutzler 1994). Furthermore, the unit absorption 
depth ($\sim 1 \times 10^{20} {\rm~cm^{-2}}$) in the band 
is substantially shorter than in the higher energy bands. Thus the background 
intensity and distribution, even at high Galactic latitudes, is sensitive 
to both the uncertainty in the \IRAS map as an X-ray-absorbing medium 
tracer and the clumpiness of the medium. Accounting for these effects
would tend to increase the corona contribution in the 1/4~keV band. 

 	In conclusion, the simple model presented here, though 
admittedly simplistic, describes reasonably well the global X-ray background 
distribution at high Galactic latitudes ($\gsim 20^\circ$) in the 0.5-2~keV 
range, and appears to be consistent with independent measurements of 
hot gas beyond the Local Bubble. The model can be improved 
when better data products (e.g., source-removed high 
resolution \ROSAT maps and \COBE DIRBE calibrated \IRAS maps ---
Snowden et al. 1997a) become available.

The author is grateful to S. L. Snowden for preprints,
W. T. Reach for information about the \IRAS and {\sl COBE} DIRBE data,
and the conference organizers for the invitation to give this 
talk. This work is supported partly by NASA under the grant NAG 5-2716.

\end{document}